\documentclass[prb,twocolumn,superscriptaddress,floats,showpacs]{revtex4}

\usepackage{graphicx}
\usepackage{bm}
\usepackage{subfigure}

\begin{document}
\draft

\title{Fidelity in topological superconductors with end Majorana fermions}

\author{Zhi Wang}

\affiliation{State Key Laboratory of Optoelectronic Materials and Technologies, School of Physics and Engineering, Sun Yat-sen University, Guangzhou 510275, China}

\author{Qi-Feng Liang}
\affiliation{Department of Physics, Shaoxing University, Shaoxing 312000, China}
\affiliation{International Center for Materials Nanoarchitectonics (WPI-MANA)\\
National Institute for Materials Science, Tsukuba 305-0044}
\author{Dao-Xin Yao}
\email{yaodaox@mail.sysu.edu.cn}
\affiliation{State Key Laboratory of Optoelectronic Materials and Technologies, School of Physics and Engineering, Sun Yat-sen University, Guangzhou 510275, China}

\date{\today}

\begin{abstract}
Fidelity and fidelity susceptibility are introduced to investigate the topological superconductors with end Majorana fermions. A general formalism is established to calculate the fidelity and fidelity susceptibility by solving Bogoliubov-de Gennes equations. It is shown that the fidelity susceptibility manifest itself as a peak at the topological quantum phase transition point for homogeneous Kitaev wire, thus serves as a valid indicator for the topological quantum phase transition which signals the appearance of Majorana fermions.
The effect of disorders is investigated within this formalism. We consider three disordered systems and observe fidelity susceptibility peak in all of them. By analyzing the susceptibility peak, we notice that the local potential disorders and the hopping disorders can shift the phase transition point, while off-diagonal disorders have no obvious influence. Our results confirm that the existence of topological quantum phase transition is robust to these disorders, while the behavior of the phase transition might be influenced by disorders.
\end{abstract}

\maketitle

\section{Introduction}
Majorana fermions (MFs), fermions which are their own anti-particles, are under intense investigation recently. MFs have been predicted as zero-energy bound states in spinless p-wave superconductors\cite{kitaev,read}, known as topological superconductors. In these topological superconductors, two MFs form one topological qubit which resists to local electro-magnetic disturbance, thus might be useful for long-time storage of quantum information\cite{kitaev,ivanov,kanermp}. Moreover,
braiding of these MFs rotates the topological qubits they form \cite{ivanov,kanermp,alicea}, thus may constitute as one of the cornerstones of quantum computation\cite{kanermp,kitaev2,sarmarmp,jiang,bonderson}.

MFs were originally proposed in pure theoretical toy models \cite{kitaev,read}. However, recent progress in spin-orbit coupling research makes it possible to be realized in hybrid systems\cite{zhangrmp}, such as superconductor-topological insulator interface\cite{fu}, or semiconductor-superconductor heterostructure \cite{sau,lutchyn,oreg,beenakker,alicea2}.
Among these systems, one-dimensional spin-orbit coupling nanowire in proximity to conventional superconductor is of particular interest, due to its relatively simple structure\cite{lutchyn,oreg}. It was predicted that there is a topological quantum phase transition (TQPT) in the system when proper Zeeman field is applied, and zero energy end MFs appear in the topological non-trial phase.
Recently, a
device with required condition has been fabricated\cite{kouwenhoven}, and the differential conductance was measured under various strength of Zeeman splitting. A robust zero-bias peak has been reported, in agreement with the resonant Andreev effect which was proposed to signify the MFs\cite{lee}. Moreover, the Zeeman energy in the device was manipulated through rotating the applied magnetic field, and the zero-bias peak was discovered under the exact condition predicted by theory.
Later on, subsequent experiments have been conducted on similar devices, and their results validated the previous report\cite{rokhinson,heiblum}.

The realization of MFs seems very optimistic with these experimental progresses. However, there are
some issues need to be fixed before the MFs can be completely pinned down. In the previous experimental reports\cite{kouwenhoven}, the extracted superconducting energy gap does not close at the TQPT point where the zero-bias peak appears, moreover, the height of the zero-bias peak is much lower than the quantized conductance $h/2e$ predicted by the resonant Andreev effect \cite{lee}. Because of these discrepancies, there are arguments on the proper interpretation to the recent experimental results.
Some provide theoretical explanations for these two issues, and try to certify the existence of MFs in the system\cite{prada,stanescu2012,law2012}. Others point out that the appearance of zero bias peak is due to non-topological mechanisms, such as disorders\cite{liu,altland,beenakker_weak}, edge confinement\cite{kells}, or Kondo effect\cite{silvano}.
While the second issue can be attributed to finite temperature effect, the first issue is rather difficult to understand. Traditionally the energy gap which protects the topology of system should close at the TQPT point, where the spectrum is reconstructed and the zero energy MFs may appear. This discrepancy indicates that the TQPT in the topological superconductors is an important problem, and might be crucial to understand the zero-bias peak. Therefore, more
investigations on the TQPT are necessary to understand the appearance of zero bias peak and its robustness.

One established method to investigate quantum phase transition is the fidelity approach\cite{gureview}. Fidelity is a measure of the difference between two quantum states. Intuitively, it is a natural marker for quantum phase transition, since it should presents a drop at the phase transition point where a dramatic reconstruction of the quantum states happens. In application, the second derivative of the fidelity, {\it i.e.} fidelity susceptibility\cite{gufidelity}, would be a better choice for application. It should manifest itself as a sharp peak at the quantum phase transition point, where the peak height scales with the system size. Fidelity and fidelity susceptibility can describe quantum phase transition without assuming \textit{a priori} knowledge of the system, thus are particularly suitable for dealing with TQPT, where the traditional symmetry-breaking formalism does not work\cite{wangfidelity}.

In this work, we apply the fidelity and fidelity susceptibility to study the zero temperature TQPT in one dimensional Kitaev wire, which supports end Majorana fermions at the topological non-trivial phase.
We establish a scheme to calculate the fidelity and fidelity susceptibility by solving the Bogoliubov-de Gennes (BdG) equations.
We confirm the validity of this scheme by comparing the results from BdG equations in open boundary system and the analytic results in periodic boundary system. In both systems, the fidelity susceptibility manifest itself as a peak at the phase transition point predicted from the topological arguement\cite{kitaev}, thus serves as a valid marker for the TQPT.
Then we introduce this scheme to investigate the disordered systems, where no information for the quantum phase transition has been drawn analytically. We consider several simple disorders, such as the local potential disorder, the hopping disorder, and the off-diagonal disorder. We find  that the fidelity susceptibility always manifest itself as a peak for all disordered systems under consideration, implying that the existence of the TQPT is robust to these disorders.
This result agrees with previous topological arguments, and supports the conclusion that the MFs are robust to disorders.
After analyzing the position and the height of the peak in fidelity susceptibility, we are able to provide detailed information for the TQPT in disordered system besides its existence. It is found that
the local potential disorders and the hopping disorders shift the phase transition point obviously, while the off-diagonal disorders have no obvious influence. We also perform scaling analysis on the TQPT, and find that the quadratic scaling behavior for homogeneous system is preserved in local potential disordered system and off-diagonal disordered system, while the scaling behavior of the hopping disordered system significantly deviates from the simple quadratic form. From these results, we are able to tell that the hopping disorder has a strong influence on the TQPT, while the off-diagonal disorder has little influence on the TQPT.
Our work illustrate that the fidelity approach is very useful to find the TQPT in topological superconductors, thus can provide an evidence for the existence of MFs, since MFs are always accompanied by a TQPT which is absent for accidental zero energy excitations. For all theoretical systems where zero energy excitations appear and argued to be MFs, fidelity approach can be applied and provides a useful check by detecting the TQPT in the system. From this point of view, our study can help to resolve the debate over the mechanism behind the zero bias peaks discovered recently\cite{kouwenhoven}.

This paper is organized as follows: The model and the analytic solution for the periodic boundary condition are discussed in Sec. II. The scheme for calculating fidelity and fidelity susceptibility by solving BdG equations, and its application to open boundary condition are presented in Sec. III. The disordered systems are solved in Sec. IV. Finally, we give a summary in Sec. V.

\section{Model and analytic solution}
The topological superconductor fabricated recently is a hybrid-system\cite{kouwenhoven}, where a spin-orbit coupling nanowire under a Zeeman field is in proximity to a conventional s-wave superconductor. A detailed model for this device should involve many factors such as Zeeman energy, spin-orbit coupling, charging energy etc. However, its key physics can be captured by the one dimensional Kitaev wire\cite{kitaev}, which is a one dimensional lattice model for a spinless superconductor,
\begin{eqnarray}
H=&& - t \sum^N_{j=1}  (c_j^\dagger c_{j+1}+c_{j+1}^\dagger c_j)
-\mu\sum^N_{j=1}  (c_j^\dagger c_j-{\textstyle\frac{1}{2}})\nonumber\\\
&&+\sum^N_{j=1}  ( \Delta c_jc_{j+1}+\Delta^* c_{j+1}^\dagger c_j^\dagger ),
\end{eqnarray}
where $c_j^\dagger$ is the electron creation operator, $t$ is the hopping integral, $\mu$ is the chemical potential, $\Delta$ is the superconducting gap, and N is the lattice size.
It has been established by Kitaev that a topological non-trivial state with end MFs appears in this model\cite{kitaev}, and the TQPT appears at $\mu=2t$ for $|\Delta|>0$, where the energy gap of the system closes. While most research concentrated on the topological non-trivial state, only few of them have targeted on the TQPT itself. In this paper, we apply the fidelity approach to investigate the TQPT in this model.
\begin{figure}[t]
\begin{center}
\includegraphics[clip=true,width=1\columnwidth]{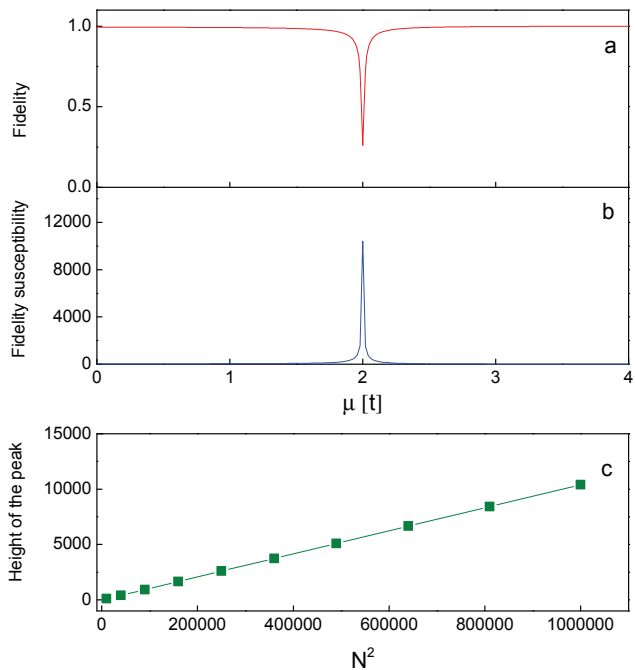}
\caption{(Color online). Fidelity (a) and Fidelity susceptibility (b) as a function of chemical potential $\mu$, and (c) the height of peak in fidelity susceptibility as a function of lattice size, where $\Delta=t$.}
\end{center}
\end{figure}
Let us first look at the the simplest case when the wire formes a ring, imposing a periodic boundary condition to the model. The
zero energy MFs vanish in this ring geometry, however, the two topological states and the TQPT between them persist.
With the periodic boundary condition, the model can be simplified via the Fourier transformation, where momentum is a good quantum number.
A pair of electrons with momentum $k$ and $-k$ will form the Cooper pair, and
a conventional BCS ground state wave-function can be obtained,
\begin{eqnarray}
|\Psi_G\rangle = \prod_{k} ({\rm u}_k + {\rm v}_k c^\dagger_k c^\dagger_{-k}) |0\rangle,
\end{eqnarray}
where quasi-particle coefficients are,
\begin{eqnarray}
{\rm u}_k = \sqrt{\frac{1}{2}+ \frac{2t \cos k + \mu}{2E_k}},{\rm v}_k = \sqrt{\frac{1}{2}- \frac{2t \cos k + \mu}{2E_k}},
\end{eqnarray}
with the energy spectrum,
\begin{eqnarray}
E_k= \sqrt{(2t \cos k + \mu)^2 + 4 |\Delta|^2 \sin^2 k}.
\end{eqnarray}
From this energy spectrum, it can be seen that the energy gap closes at $\mu=2t$, if $|\Delta|>0$. This special point where the energy gap vanishes is argued to be the TQPT point of the system\cite{kitaev}, since topology of the system is protected by the energy gap.

Choosing the chemical potential $\mu$ as the driving parameter, the fidelity between two nearby ground states is obtained straightforwardly with this wave-function,
\begin{eqnarray}
f&&= \langle \Psi_G(\mu)|\Psi_G(\mu+\delta \mu)\rangle\\\
 &&= \prod_{k} \left[{\rm u}_k(\mu){\rm u}_k(\mu+\delta \mu) + {\rm v}_k(\mu){\rm v}_k(\mu+\delta \mu) \right], \nonumber
\end{eqnarray}
and the fidelity susceptibility is the derivative of the fidelity, with an analytical form,
\begin{eqnarray}
\chi_F&&=  2 \lim_{\delta\mu\rightarrow 0} \frac{1-|\langle \Psi_G(\mu)|\Psi_G(\mu+\delta\mu)\rangle|^2}{\delta\mu^2}\nonumber
\\\
&&=\langle \frac{d\Psi_G}{d\mu}|\frac{d\Psi_G}{d\mu}\rangle -\langle \frac{d\Psi_G}{d\mu}|{\Psi_G}\rangle\langle {\Psi_G}|\frac{d\Psi_G}{d\mu}\rangle\nonumber\\\
 &&= \sum_{k} \frac{\Delta^2 \sin^2 k}{E_k^4}.
\end{eqnarray}
The fidelity susceptibility is simply determined by the energy spectrum, however, this simple expression is applicable only under the periodic boundary condition.
\begin{figure}[t]
\begin{center}
\includegraphics[clip=true,width=1\columnwidth]{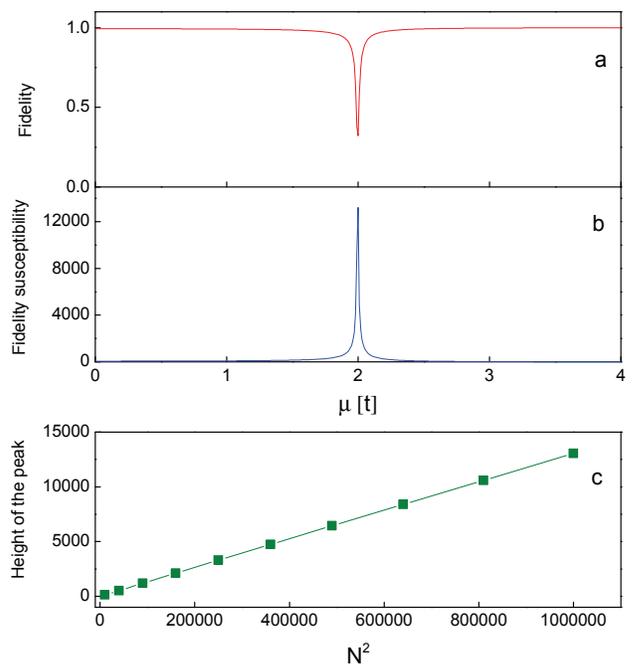}
\caption{(Color online). Same as in Fig. 1 except for open boundary condition.}
\end{center}
\end{figure}

The results of fidelity and fidelity susceptibility with a system size $N=1000$ are shown in Fig. 1a. and Fig. 1b. as a function of chemical potential $\mu$. It can be seen that the fidelity susceptibility manifests itself as a sharp peak at $\mu=2t$, and clearly signatures the TQPT point speculated in previous researches\cite{kitaev}.
Our results confirm the previous argument based on the energy spectrum analysis\cite{kitaev}, and validate the fidelity approach for marking the TQPT in this system.

We also perform a scaling analysis for the fidelity susceptibility at the TQPT point. The height of the peak is shown as a function of the lattice size $N$ in Fig. 1c. It is obvious that the height of the peak quadratically scales with the system size, and would diverge when approaching the thermodynamic limit. We notice that our result for Kitaev model should be the same to the previous results obtained in the transverse-field Ising model \cite{gureview}, since these two models are equivalent to each other in the absence of disorders.

\section{open boundary problem}
The most fascinating aspect of topological superconductor is the topologically protected end Majorana fermions, which unfortunately is missing in the ring-like structure as we discussed in the previous section. To get the zero energy MFs, we must consider a wire with ends which enforce an open boundary condition.
In this case, no analytical solution is available, due to the absence of the translational symmetry.
A conventional method to tackle this system is to solve the Bogoliubov-de Gennes (BdG) equations, diagonalize the Hamiltonian in real space and obtain the energy spectrum as well as the quasi-particle wave functions.
In order to calculate the fidelity and fidelity susceptibility, we have to construct the wave function of the system. However, a wave function in the electron representation like the one given in Eq. (2) is absent, since the pairing does not happen in momentum space. On the other hand, the ground state wave function can be constructed within the quasi-particle representation simply by filling all the negative energy quasi-particle states, similar to the picture of fermi sea.

In the following, we utilize this formalism and find that the fidelity and fidelity susceptibility naturally arises.
The system we study here is a finite size Kitaev wire with N sites. The energy spectrum $E_n$ and the quasi-particle wave-function
$\psi_{n}(j) = ({\rm u}_{nj},{\rm v}_{nj})$ can be obtained by numerically solving the BdG equations, where n runs from $1$ to $2N$ denoting the quantum number of the energy level, and $j$ runs from $1$ to $N$ denoting the site number.
As we have discussed, the ground state wave function should be the combination of the wave functions of all negative energy quasiparticle states, which can be expressed as a Slater determinate,
\begin{eqnarray}
|\Psi(j_1,\cdots,j_n) \rangle=
 \frac{1}{\sqrt{n!}}\left|
 \begin{array}{ccc}
\psi_1({j_1})&\cdots&\psi_1({j_n})\\
 \vdots&\ddots&\vdots\\
  \psi_n({j_1})&\cdots&\psi_n({j_n})
\end{array}
  \right|,
\end{eqnarray}
where $n$ denotes the number of quasi-particle states with negative energy. For the open boundary Kiatev wire, there are zero energy MF bound states at both ends in topological non-trivial phase, making the ground state doubly degenerate. In order to avoid complexity, we added quasi-particle wave function of these zero energy states into Eq. (7), making $n=N+1$ in our calculation.

Inserting this wave function into the definition of fidelity Eq. (5), and taking advantage of a simplification for calculating fidelity in multi-particle fermion system,
\begin{widetext}
\begin{eqnarray}
f &&=\langle \Psi_{\mu}|\Psi_{\mu + \delta \mu}\rangle  = \frac{1}{n!}\int dx_1\cdots dx_n
 \left|
 \begin{array}{ccc}
\psi^*_1({x_1})&\cdots&\psi^*_1({x_n})\\
 \vdots&\ddots&\vdots\\
  \psi^*_n({x_1})&\cdots&\psi^*_n({x_n})
\end{array}
  \right|_{\mu} \cdot \left|
 \begin{array}{ccc}
\psi_1({x_1})&\cdots&\psi_1({x_n})\\
 \vdots&\ddots&\vdots\\
  \psi_n({x_1})&\cdots&\psi_n({x_n})
\end{array}
  \right|_{\mu+\delta \mu}
 \nonumber \\\ &&= \frac{1}{n!}\int dx_1\cdots dx_n
 \left|\left(
 \begin{array}{ccc}
\psi^*_1({x_1})&\cdots&\psi^*_1({x_n})\\
 \vdots&\ddots&\vdots\\
  \psi^*_n({x_1})&\cdots&\psi^*_n({x_n})
\end{array}
  \right)_{\mu} \cdot \left(
 \begin{array}{ccc}
\psi_1({x_1})&\cdots&\psi_n({x_1})\\
 \vdots&\ddots&\vdots\\
  \psi_1({x_n})&\cdots&\psi_n({x_n})
\end{array}
  \right)_{\mu+\delta \mu}\right|
= \left|
 \begin{array}{ccc}
P_{11}&\cdots&P_{1n}\\
 \vdots&\ddots&\vdots\\
  P_{n1}&\cdots&P_{nn}
\end{array}
  \right|,
\end{eqnarray}
\end{widetext}
where $\Psi_{\mu}$ indicates the wave-function under chemical potential $\mu$, and $P_{lm} = \int dx \psi^*_{
\mu,l}({x})\psi_{\mu+\delta
\mu,m}({x})$.
With this expression for fidelity, the fidelity susceptibility can be obtained through the definition
\begin{eqnarray}
\chi_F&&=  2 \lim_{\delta\mu\rightarrow 0} \frac{1-|\langle \Psi_\mu|\Psi_{\mu+\delta\mu} \rangle|^2}{\delta\mu^2},
\end{eqnarray}
where the differentiation is calculated numerically.
The result of the fidelity and fidelity susceptibility are shown in Fig. 2a and Fig. 2b, and the scaling behavior for the peak of the fidelity susceptibility is shown in Fig. 2c, with all physical parameters the same as in the previous section.
\begin{figure}[tb]
\begin{center}
\includegraphics[clip=true, width=1\columnwidth]{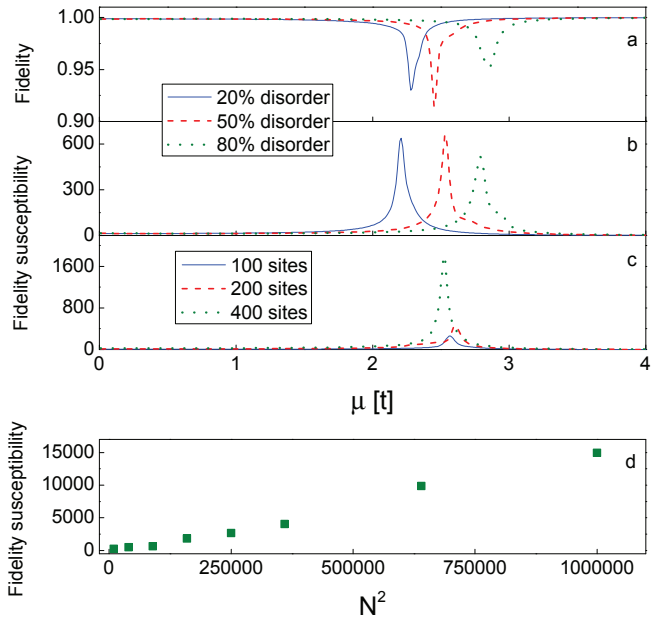}
\caption{(Color online). Fidelity (a) and Fidelity (a) susceptibility as a function of chemical potential $\mu$ with disorder concentration of $20$$\%$ (solid line), $50$$\%$ (dashed line), and 80$\%$ (dotted line), where lattice size $N=200$. (c) The fidelity susceptibility of with lattice size $N= 100$, $200$, and $400$, where the disorder concentration is $50\%$. (d) The height of the peak in fidelity susceptibility as a function of the lattice size $N$, with $50\%$ of disorders.}
\end{center}
\end{figure}
We find that the fidelity susceptibility manifest itself as a sharp peak at the TQPT point, showing that it is capable of signifying the TQPT in the wire. The height of the peak is quadratically scaling with the system size,
similar to the results discovered in Fig. 1c which comes from the analytic solutions in periodic boundary condition. Considering that the boundary condition should be irrelevant to the phase transition for large system, this near identical result illustrates that our scheme of calculating fidelity by solving the BdG equations is a valid procedure, and would be helpful to investigate the TQPT in the superconducting systems.

\section{disorder effects}
After establishing the fidelity approach to the TQPT by solving the BdG equations, we are now ready to deal with more realistic and complicated disorder systems. The role of disorders in experimental device is still under debate. In traditional believe, its effect should be very weak. However some recent theories argue that disorder itself might be the origin of the zero-bias peak, making the zero-bias peak irrelevant to the MFs\cite{liu,altland,beenakker_weak}.
In this section, we investigate the effects of the disorders on the TQPT based on the scheme we built in last section. Since TQPT is necessary for the appearance of MFs, our study will also provide the information on the robustness of the MFs.

We study some examples to show the
power of fidelity approach to investigate the TQPT in disordered system, and illustrate
the influence of several typical disorders on the TQPT.
The most typical disorder is a local potential disorder, which modulates the local chemical potential with the Hamiltonian,
\begin{eqnarray}
H_1= V_1 c_i^\dagger c_i
\end{eqnarray}
where $i$ indicates the site with the impurity, and $V_1$ indicates the local potential.
We consider a random distribution of the same kind of disorders on the wire, and investigate the behavior of the TQPT for different disorder concentrations. The two mathematical limits of one hundred percent of disorders and zero percent of disorders are trivial and well understood theoretically. In the first limit, an extra potential should be enforced to every site, which is equivalent to shifting the total chemical potential from $\mu$ to $\mu-V_1$, thus the phase transition point should be shifted from $\mu=2t$ to $\mu=2t+V_1$. In the second limit, there is no disorder in the system, and the phase transition point should be the same as in the previous section. However, the properties of the TQPT, even its existence, for a intermediate concentration has to be investigated within fidelity approach.
\begin{figure}[t]
\begin{center}
\includegraphics[clip=true, width=1\columnwidth]{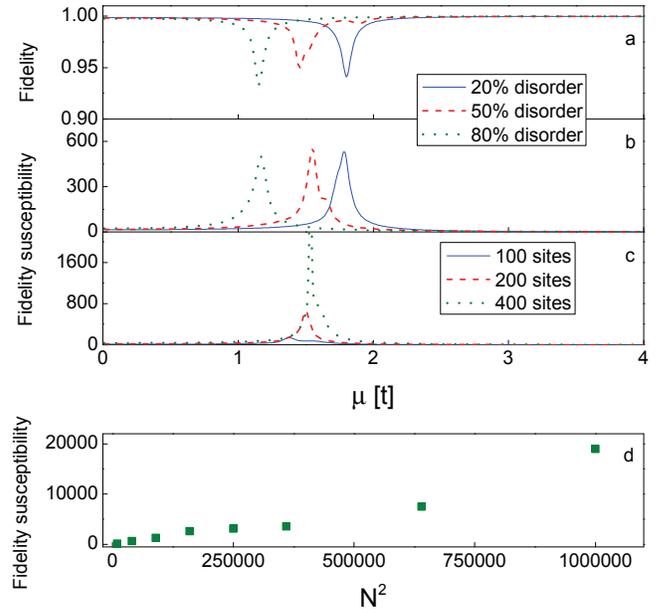}
\caption{(Color online). Same as in Fig. 3 except for hopping disorders.}
\end{center}
\end{figure}
The results of the fidelity and fidelity susceptibility with $20$, $50$ and $80$ percent of potential disorders are shown in Fig 3a, and Fig. 3b, with the disorder strength $V_1=t$. We can see that for all disorder concentrations, fidelity susceptibility manifest itself as a peak at different places.
It is a clear evidence that the TQPT is robust to disorder, while the phase transition point shifts when changing the disorder concentration.
In order to further confirm the robustness of the TQPT, we present the fidelity susceptibility
for different lattice size of $100$, $200$, and $400$ in Fig. 3c, with $50$ percent of disorders.
It can be seen that the sharp peak in fidelity susceptibility appears for all lattice sizes, while the peaks are located at almost the same positions. We argue that the small deviation of peak positions might come from different specific disorder configurations, and reflect the finite size effect. We present the scaling behavior of the peak height in Fig. 3d, and find it roughly scales quadratically with the lattice size.
However, we have to mention that this scaling behavior is very inaccurate, since the peak height still depends on the specific disorder configurations within the lattice size we have investigated. We have performed more calculations for different lattice sizes and different disorder strengths, as well as ensemble averages for the disorder, and find that the peak in fidelity susceptibility is very robust. These results clearly suggest that the TQPT, as well as the appearance of MFs, will not be suppressed by the local potential disorders.

\begin{figure}[tb]
\begin{center}
\includegraphics[clip=true, width=1\columnwidth]{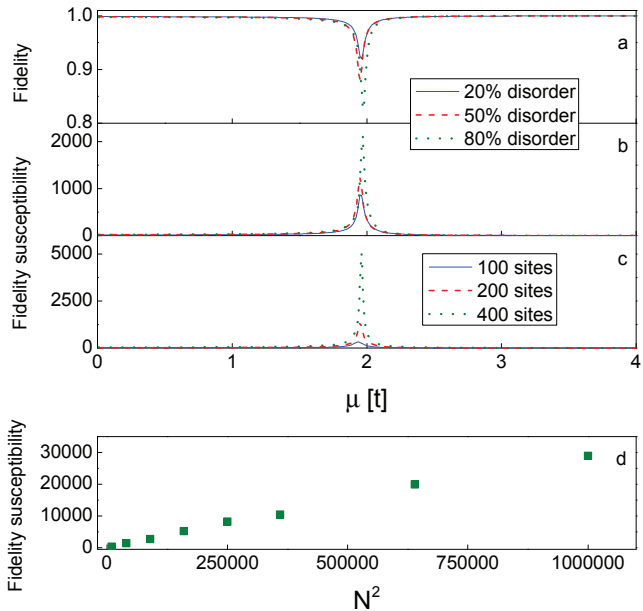}
\caption{(Color online). Same as Fig. 3 except for off-diagonal disorders.}
\end{center}
\end{figure}
The second type of disorder we investigate is the hopping disorder, which modulates the local hopping integral between sites with the Hamiltonian,
\begin{eqnarray}
H_2=  V_2 c_i^\dagger c_{i+1} +h. c.
\end{eqnarray}
where the disorder is located at the bond between site $i$ and $i+1$, with $V_2$ the disorder strength. Similar to the local potential disorders, the limiting case of the disorder concentration is trivial. For one hundred percent of disorders, the phase transition point is simply shifted to $\mu = 2(t-V_2)$. We present the results of fidelity and fidelity susceptibility with $20$, $50$, and $80$ percent of hopping disorders in Fig. 4a and Fig. 4b , where the disorder strength is $V_2=0.5t$. We can see that the influence of hopping disorder is very similar to the local potential disorder, that is, The TQPT is robust while the phase transition point is shifted.
The fidelity susceptibility for different system size is shown in Fig. 4c, and it is found that the peak always appears indicating the existence of a TQPT.
The scaling behavior of the peak height is presented in Fig. 4d, and it is found that the quadratic scaling for homogeneous system barely exists anymore. These results illustrate that the hopping disorder preserves the TQPT in the system, however, it has a larger influence on the detailed behavior of the TQPT than the local potential disorders, which can be seen from the scaling behavior of height of the peak in fidelity susceptibility.

Finally, we are going to investigate the off-diagonal disorder, which has been extensively studied in high temperature superconductors\cite{wangcuprate}. The off-diagonal disorder can be understood as a static fluctuation of the amplitude of superconducting gap. In a mean-field level, this disorder can be described with the Hamiltonian,
\begin{eqnarray}
H_{3}=  -V_3 c_i^\dagger c_{i+1}^\dagger +h. c.
\end{eqnarray}
where the disorder is located at the bond between sites $i$ and $i+1$, with $V_3$ the modulation to the local superconducting gap amplitude. For this off-diagonal disorder, the two limit of the disorder concentration is very similar, since the TQPT should be irrelevant to detailed superconducting gap values as long as it is non-zero\cite{kitaev}. In this case, we would expect that the TQPT is less influenced by these disorders. In Fig. 5a and Fig. 5b, we show the results of fidelity and fidelity susceptibility for $20$, $50$, and $80$ percent of off-diagonal disorders, with a disorder strength $V_3=0.5t$. As we expected, we find that the fidelity susceptibility manifest a sharp peak almost at the same position as in the homogeneous system, indicating a minimal influence to the TQPT. We also present the fidelity susceptibility for different system sizes in Fig. 5c with $50$ percent of disorder, and find that the peak always appears.
The scaling behavior of the peak height is shown in Fig. 5d, and it is found that the height of the peak is scaling quadratically with the system size, which is very close to the result for homogeneous system in the last section. These results illustrate a minimal influence of these off-diagonal disorders, and confirmed that they are irrelevant to the TQPT and the MFs.

We have studied three types of the disorders in this section and found that all of them preserve the TQPT. While the on-site potential disorder and hopping disorder shift the phase transition point and modulate the scaling behavior of peaks in fidelity susceptibility, the off-diagonal disorder has a minimal influence to the TQPT. Our results show that the TQPT, together with the topological non-trivial state and the end MFs, is robust to these disorders.

\section{Summary and discussion}

In summary, we have investigated the topological quantum phase transition in the Kitaev model within the fidelity approach. A formalism to calculate the fidelity and fidelity susceptibility by solving the BdG equations has been established and its validity is confirmed by comparing to the analytic results in homogeneous system. This formalism is then applied to investigate the disordered systems, where no analytic solution is available. Three typical disordered systems are investigated, and it is found that the disorders preserve the TQPT based on the appearance of the peak in fidelity susceptibility. From the peak positions, we find that the diagonal disorders shift the phase transition point clearly, while the off-diagonal disorders has no obvious impact on the peak position.

\section{Acknowledgement}
We thank S. P. Kou, X. Hu, K. T. Law, T. K. Ng and L. Fu for very helpful discussions. This work was supported by Fundamental Research Funds for the Central Universities of China under Grant No. 30000-1188131. Q.-F.Liang is supported by NSFC-10904092.
D.X.Y. is supported by the National Basic Research Program of China (2012CB821400), NSFC-11074310, NSFC-11275279, Specialized Research Fund for the Doctoral Program of Higher Education (20110171110026), Fundamental Research Funds for the Central Universities of China, and NCET-11-0547.

\end{document}